# Control of a Bucket-Wheel for Surface Mining of Asteroids and Small-Bodies

Ravi teja Nallapu,[*] Erik Asphaug,[†] and Jekan Thangavelautham[‡]

Near Earth Asteroids (NEAs) are thought to contain a wealth of resources, including water, iron, titanium, nickel, platinum and silicates. Future space missions that can exploit these resources by performing In-Situ Resource Utilization (ISRU) gain substantial benefit in terms of range, payload capacity and mission flexibility. Compared to the Moon or Mars, the milligravity on some asteroids demands a fraction of the energy for digging and accessing hydrated regolith just below the surface. However, asteroids and small-bodies, because of their low gravity present a major challenge in landing, surface excavation and resource capture. These challenges have resulted in adoption of a "touch and go techniques", like the upcoming Osiris-rex sample-return mission. Previous asteroid excavation efforts have focused on discrete capture events (an extension of sampling technology) or whole-asteroid capture and processing. This paper analyzes the control of a bucket-wheel design for asteroid or small-body excavation. Our study focuses on system design of two counter rotating bucket-wheels that are attached to a hovering spacecraft. Regolith is excavated and heated to $1000^{\circ}$C to extract water. The water in turn is electrolyzed to produce hydrogen and oxygen for rocket fuel. We analyze control techniques to maximize traction of the bucket-wheels on the asteroid surface and minimize lift-off the surface, together with methods to dig deeper into the asteroid surface. Our studies combine analytical models, with simulation and hardware testing. For initial evaluation of material-spacecraft dynamics and mechanics, we assume lunar-like regolith for bulk density, particle size and cohesion. Our early studies point towards a promising pathway towards refinement of this technology for demonstration aboard a future space mission.

**INTRODUCTION**

Asteroids are known to contain large reserves of water, iron, titanium, nickel, platinum, palladium and silicates[1]. In-Situ Resource Utilization (ISRU)[2] aims at extracting these resources on-site, which if successful, can improve our current space exploration capabilities. While there has been progress reported in ISRU research for applications on the lunar surfaces[3] and Mars[4], the low-gravity environments on asteroids, on the other hand, presents additional challenges in performing ISRU[5]. One of the major challenges includes lift-off due to excavator thrusting off the small body.

---

[*] PhD Student, Space and Terrestrial Robotic Exploration Laboratory, Arizona State University, 781 E. Terrace Mall, Tempe, AZ.
[†] Professor and Ronald Greeley Chair of Planetary Science, Space and Terrestrial Robotic Exploration Laboratory, Arizona State University, 781 E. Terrace Mall, Tempe, AZ
[‡] Assistant Professor, Space and Terrestrial Robotic Exploration Laboratory, Arizona State University, 781 E. Terrace Mall, Tempe, AZ

In this paper we focus on use of bucket-wheels to perform excavation. Bucket-wheels have been identified as one of the best tools for excavation in low-gravity environments, beating out bulldozers and front-loaders[10,13]. This paper addresses the problem of excavator lift-off which is the predominant concern on very low-gravity environments; by considering a dual bucket wheel excavator on the surface of Martian moon Phobos. Phobos as shown in Figure 1 has an escape velocity of about 11.1 m/s. Hence if any of the excavation mechanisms experience a force that can impart this speed, the mechanism would be thrown off the surface of Phobos.

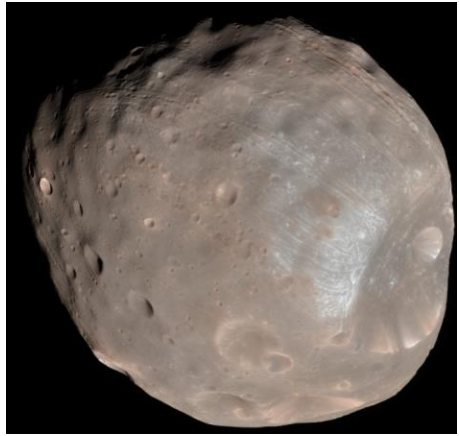

**Figure 1. Martian moon Phobos is a promising target for ISRU.**

In this paper, we present a bucketwheel excavator system that has two counter rotating bucket wheels, like RASSOR developed by NASA KSC[6], and can apply the necessary vertical plunge force. The counter rotation ensures that no horizontal velocities are imparted, and the vertical plungers ensure the compensation for lift-off due to any reactive-excavation forces.

The paper starts with introduction to the bucket-wheel system design followed by optimization of its size, operating parameters and power source [7,8]. We then present dynamical and control models of the system and a simulation of the proposed mechanism, followed by preliminary results and conclusions.

**BUCKET WHEEL SYSTEM DESIGN**

The design of a bucketwheel is a multi-disciplinary problem. Several forces, geometry, and performance requirements are considered. To understand the role of an excavator, a diagram of the full ISRU system is shown in Figure 2.

The excavation system is powered by solar insolation. The solar panels generate the power based on the incident radiation, which is used to charge an onboard system battery that powers various subsystems.

The ISRU unit is tasked with excavating regolith, extracting water by heating regolith and electrolyzing hydrogen and oxygen from the water. The hydrogen and oxygen would be used as propellant for rockets on their journey to earth or to power fuel cells power supplies during eclipse[14,15,16].

A design of a bucket-wheel and ISRU system was provided by Nallapu et al.[8] (Figure 3) where the geometric parameters of the bucket wheel, including number of buckets and their dimensions were obtained by constraining the total operating power to 10 kW.

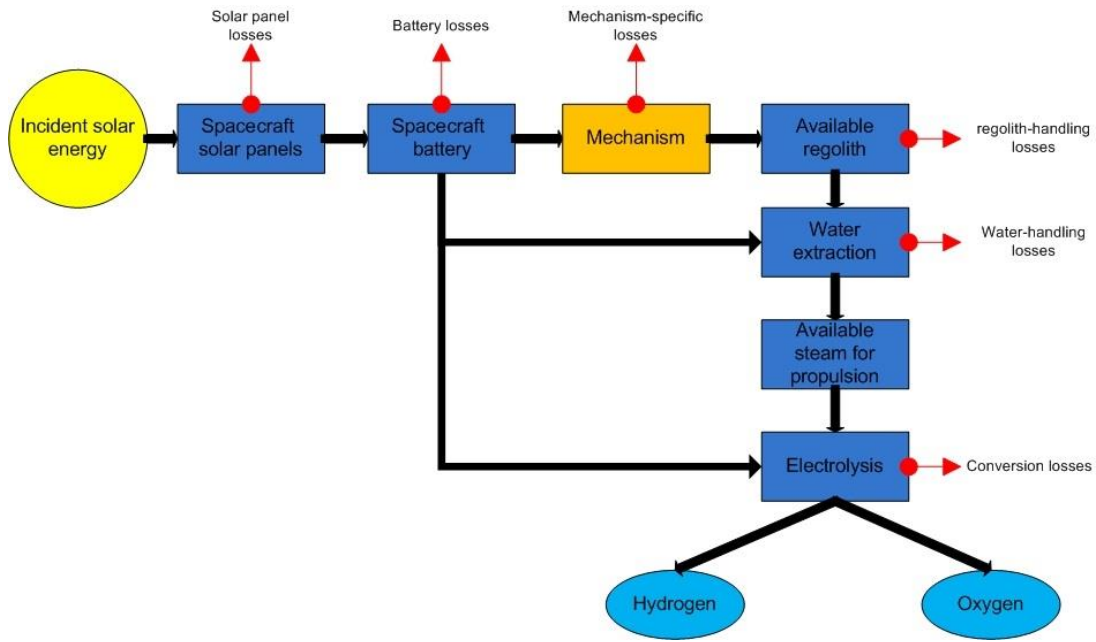

**Figure 2. Systems diagram for the proposed ISRU system**

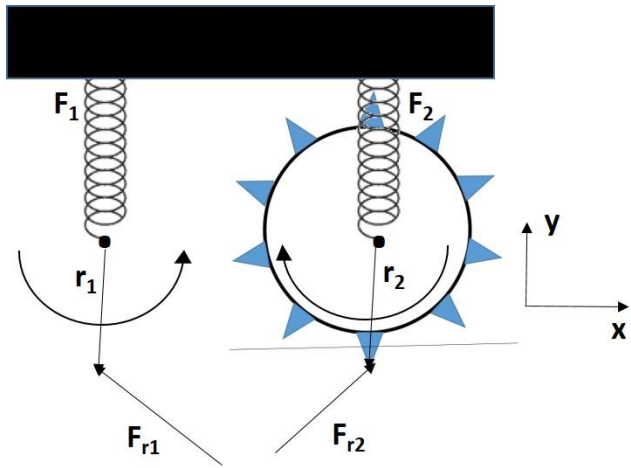

**Figure 3. Proposed excavator with counter rotating bucket wheels**

This proposed unit consists of 2 counter rotating bucket-wheels that cancel out horizontal motion. The unit is also capable of compensating for any lift-off vertical forces by exerting a radial thrust as shown in Figure 3.

## SYSTEM DYNAMICS

### Excavation dynamics

Consider a bucket wheel excavator, plunging the bucket into the regolith as shown in Figure 4. Let the wheel be applied torque, $\tau$, and be spinning with a speed, $\omega$. During the excavation, the wheel experiences the resistive forces from the regolith, $F_{reg}$, normal to the cutting surface. Let $F_{sup}$, a force applied to counter-act any lift-off experienced.

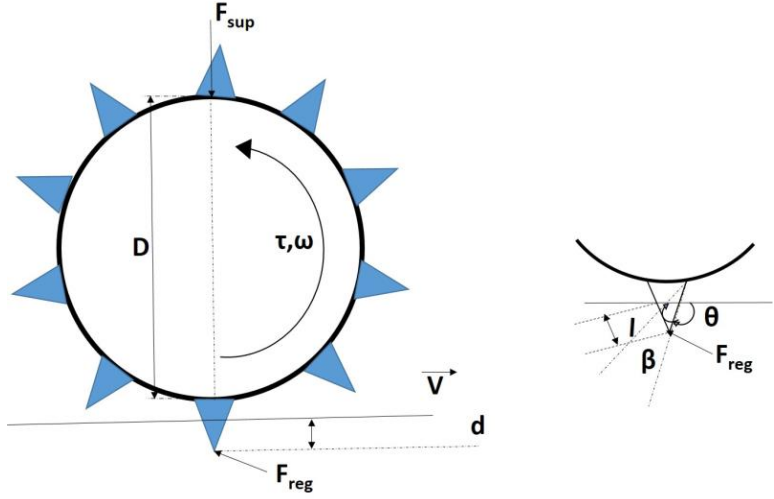

**Figure 4. Excavation modelling**

The resistive regolith forces are modeled based on a Luth-Wismer model[9]. The Luth-Wismer model separates the regolith forces as a non-cohesive force (sand) and pure cohesive (clay) forces. Skonieczny et al.[10] formulates these forces as:

$$F_{sand} = \rho g w l^{1.5} \beta^{1.73} \sqrt{d} \left(\frac{d}{l \sin \beta}\right)^{0.77} \left[1.05 \left(\frac{d}{w}\right)^{1.11} + 1.26 \frac{v^2}{gl} + 3.91\right] \quad (1)$$

and,

$$F_{clay} = \rho g w l^{1.5} \beta^{1.15} \sqrt{d} \left(\frac{d}{l \sin \beta}\right)^{1.21} \left[\left(\frac{11.5c}{\rho g d}\right)^{1.21} \left(\frac{2v}{3w}\right)^{0.121} \left(0.055 \left(\frac{d}{w}\right)^{0.78} + 0.065\right) + 0.64 \left(\frac{v^2}{gl}\right)\right] \quad (2)$$

The force acts normal to the cutting surface, and the total force is given by:

$$F_{sand} + F_{clay} = F_{reg} \quad (3)$$

**Equations of motion**

This method of separating the forces along the horizontal and vertical axis is similar in procedure used by Park[11] and Xiao et al.[12]. The equations of motion for the system shown in Figure 3 are therefore given by:

$$m\ddot{x} = -F_{r1}\sin(\beta_1 + \theta_1) + F_{r2}\sin(\beta_2 + \theta_2) \quad (4)$$

$$m\ddot{y} = F_{r1}\cos(\beta_1 + \theta_1) + F_{r2}\cos(\beta_2 + \theta_2) - mg - F_1 - F_2 \quad (5)$$

$$I\dot{\omega}_1 = \tau_1 + r_1 \times F_{r1} \quad (6)$$

$$I\dot{\omega}_2 = \tau_2 + r_2 \times F_{r2} \quad (7)$$

$$\dot{\theta}_1 = \omega_1 t \quad (8)$$

$$\dot{\theta}_2 = \omega_2 t \quad (9)$$

where $F_1$ and $F_2$ are the forces applied to counteract any lift-off forces on wheels 1 and 2 respectively. The forces $F_{r1}$ and $F_{r2}$ are the resistive forces on regolith acting on wheels 1 and 2 respec-tively, and whose magnitude is calculated by Equation 3. Additionally, a random disturbance force in the range [0, $F_{reg}/2$] was added to the magnitudes of $F_{r1}$ and $F_{r2}$ to account for any unmodeled dynamics. Finally, $\tau_1$ and $\tau_2$ represent the control torques exerted on the wheels 1 and 2 respectively.

**SIMULATION**

The design of the wheel as mentioned earlier was to maintain the total operational power under 10 kW. This required that the wheels' spin at about 3 RPM[8], which is the value of $\omega_{des}$ in Equations 11 and 12. Additionally, the translator motion needs to be confined vertically downwards, and no horizontal motion was desired. To realize this, a proportional-derivative (PD) controller was applied to the vertical forces, and wheel torques as given by Equations 10-12. The control law given by Equation 10 is only applied when y>0. The gains $K_y$, $K_{vy}$, $K_1$, $K_2$, and $K_x$ are chosen by repetitively running simulations and picking the values that give satisfactory performance.

$$F_1 + F_2 = -K_y y - K_{vy} V_y \tag{10}$$

$$\tau_1 = -K_1(\omega_1 - \omega_{des}) - K_x x \tag{11}$$

$$\tau_2 = -K_2(\omega_2 + \omega_{des}) - K_x x \tag{12}$$

Equations 4 through 5 were propagated with an ODE45[13] solver in MATLAB with the following parameters shown in Table 1.

**Table 1. Simulation Parameters.**

| Simulation Parameter | Value |
|---|---|
| # Wheels | 2 |
| # Buckets per wheel | 24 |
| Mass of wheel (kg) | 5 |
| D (m) | 0.622 |
| W (m) | 6.31E-02 |
| $\beta_1$ (degrees) | 10 |
| $\beta_2$ (degrees) | -10 |
| V (m/s) | 0.12 |
| p (kg/m$^3$) | 1880 |
| g (m/s$^2$) | 0.0057 |
| c (Pa) | 147 |
| Specific heat of the material (J/[Kg-Celsius]) | 1,430 |
| Surface temperature of the material (Celsius) | 200 |
| Water extraction temperature (Celsius) | 1,000 |
| Water content in the regolith (%) | 10 |
| $K_x$ | 1 |
| $K_y$ | 0.9 |
| $K_{vy}$ | 90,000 |
| $K_1$ | 4,000 |
| $K_2$ | 4,000 |

## RESULTS

This section presents the results of simulations described above. These simulations were run for a time length of 100 seconds with a step length of 0.1 seconds.

### Excavator Displacement

The displacement response of the excavator is shown in Figure 5. As seen here, the displacement is confined to the negative *y* direction, while the counter rotating wheels ensure that there is no horizontal displacement.

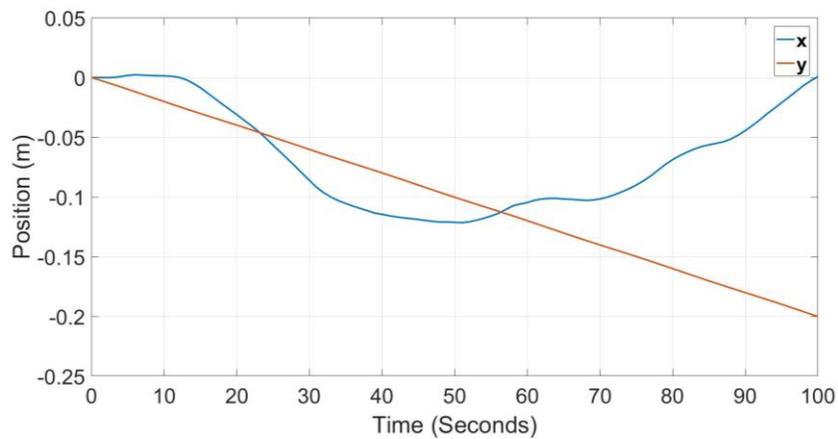

**Figure 5. Target position of bucket-wheel**

### Angular Velocity Response

Angular velocity response of the 2 bucket wheels are shown in Figure 6. As seen here the 2 wheels reach the desired speed of 3.3 RPM almost instantly while rotating in opposite directions.

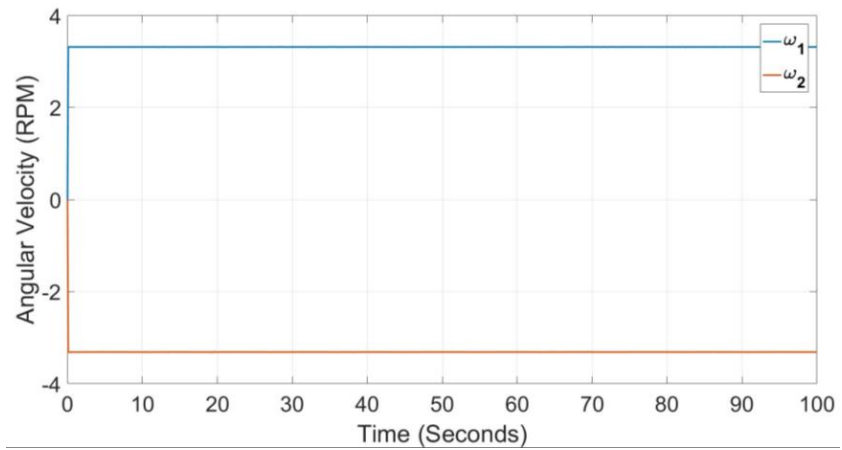

**Figure 6. Angular velocity response**

**Wheel Motion**

The rotations of the two bucket wheels are shown in Figure 7. As expected, their magnitudes are equal in opposite directions because of the counter rotations of the wheel.

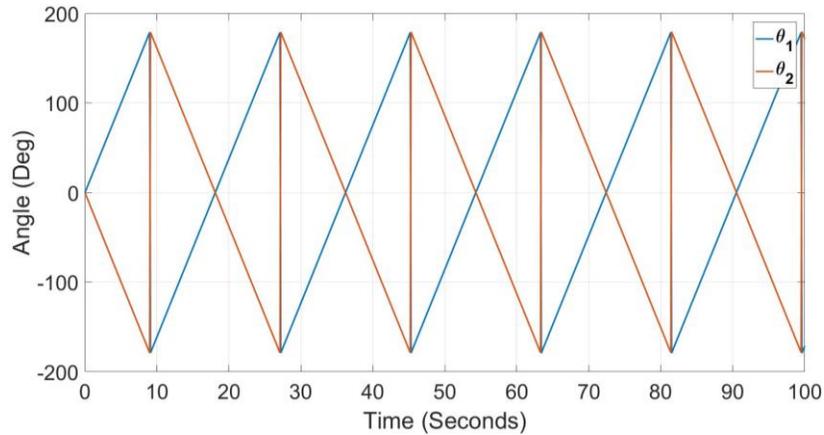

**Figure 7. Wheel rotations**

**CONCLUSIONS**

In this paper, we present the significance of ISRU in extracting resources from small-bodies such as Martian moon Phobos. Phobos is strategic location and a potential base-camp for a future human mission to Mars. This paper identifies one of the principle challenges of ISRU, namely lift-off of the excavation system in low-gravity environments such as Phobos. A dynamics and control model of a 10 kW excavator and ISRU system is presented, and its equations of motion are then propagated with computer simulations. The simulations show the proposed bucket-wheel excavator can indeed be stable on an asteroid surface, and is not susceptible to liftoff forces. Further work will be required in testing algorithms in a relevant environment and in identifying the required sensors and actuators to maintain a steady-pace of excavation on the surface of the small-body.

# REFERENCES

[1] W. F. Bottke, *Asteroids III*, University of Arizona Press, Tucson, Arizona 2002.

[2] K. Zacny, P. Chu, G. Paulsen, et al. 'Mobile in-Situ Water Extractor (MISWE) for Mars, Moon, and Asteroids in Situ Resource Utilization', *AIAA SPACE Conference*, 2012.

[3] P. Mueller, and R. H. King. 'Trade Study of Excavation Tools and Equipment for Lunar Outpost Development and ISRU', *Space Technology and Applications International Forum (STAIF 2008)*, vol. 969, 2008, pp. 237-244.

[4] G. B. Sanders, A. Paz, L. Oryshchyn, et al. 'Mars ISRU for Production of Mission Critical Consumables – Options, Recent Studies, and Current State of the Art', *AIAA SPACE 2015 Conference and Exposition*, 2015.

[5] A. Probst, G. Peytavi, B. Eissfeller, et al. 'Mission Concept Selection for an Asteroid Mining Mission', *Aircraft Engineering and Aerospace Technology*, vol. 88, no. 3, 2016, pp. 458-470.

[6] R. Mueller, R. E. Cox, T. Ebert, et al. 'Regolith Advanced Surface Systems Operations Robot (RASSOR)', *2013 IEEE Aerospace Conference*, 2013, pp. 1-12.

[7] T. Nakamura, and B. K. Smith. 'Solar Thermal System for Lunar ISRU Applications: Development and Field Operation at Mauna Kea, HI', *Proceedings of SPIE*, vol. 8124, no. 1, 2011.

[8] R. T. Nallapu, A. Thoesen, L. Garvie, E. Asphaug, J. Thangavelautham. "Optimized Bucket Wheel Design for Asteroid Excavation," *International Astronautic Congress*, 2016, Guadalajara, Mexico, IAF.

[9] H.J. Luth and R.D. Wismer, 'Performance of Plane Soil Cutting Blades in Sand'. *Journal of Terramechanics*, vol. 9, no. 4, 1973, pp. 67-67.

[10] K. Skonieczny,. 'Lightweight Robotic Excavation', *PhD Thesis*, Carnegie Mellon University, 2013.

[11] B. Park. 'Development of a Virtual Reality Excavator Simulator: A Mathematical Model of Excavator Digging and a Calculation Methodology', 2002.

[12] Thangavelautham, J., Abu El Samid, N., Grouchy, P., Earon E., Fu, T., Nagrani, N., D'Eleuterio, G.M.T., "Evolving Multirobot Excavation Controllers and Choice of Platforms Using Artificial Neural Tissue Controllers," *Proceedings of the IEEE Symposium on Computational Intelligence for Robotics and Automation*, 2009, DOI: 10.1109/CIRA.2009.542319

[13] J. Thangavelautham, K. Law, T. Fu, N. Abu El Samid, A. Smith, G. M.T. D'Eleuterio, "Autonomous Multirobot Excavation for Lunar Applications," *Robotica*, pp. 1-39, 2017.

[14] J. Thangavelautham and S. Dubowsky, "On the Catalytic Degradation in Fuel Cell Power Supplies for Long-Life Mobile Field Sensors." *Journal of Fuel Cells: Fundamental to Systems*, pp. 181-195, 2013.

[15] P. Iora, J. Thangavelautham, "Design of a mobile PEM power backup system through detailed dynamic and control analysis," *International Journal of Hydrogen Energy*, vol. 37, no. 22, pp. 17191–17202, 2012.

[16] J. Thangavelautham, D. Strawser, M. Cheung, S. Dubowsky, "Lithium Hydride Powered PEM Fuel Cells for Long-Duration Small Mobile Robotic Missions," *IEEE International Conference on Robotics and Automation (ICRA)*, St. Paul, Minnesota, 2012.